\documentclass{article}

 \usepackage[dblblindworkshop, final]{neurips_2025}

\usepackage[utf8]{inputenc} 
\usepackage[T1]{fontenc}    
\usepackage{hyperref}       
\usepackage{url}            
\usepackage{booktabs}       
\usepackage{amsfonts}       
\usepackage{nicefrac}       
\usepackage{microtype}      
\usepackage{xcolor}         
\usepackage{graphicx}
\usepackage{tikz}
\usepackage{wrapfig}
\usetikzlibrary{arrows.meta,positioning}
\usepackage{aas_macros}

\tikzset{
  block/.style={
    draw, rounded corners,
    inner sep=3pt,
    align=center,
    text width=0.25\textwidth  
  },
  arrow/.style={-Latex}
}

\title{\texttt{RUBIX}: Differentiable forward modelling of galaxy spectral data cubes for gradient-based parameter estimation}

\author{%
\href{mailto:annalena.schaible@iwr.uni-heidelberg.de}{Anna Lena Schaible$^{1,2}$}, 
\href{}{Ufuk Çakır$^3$}, 
\href{}{Tobias Buck$^{1,2}$}, 
\href{}{Harald Mack$^1$},
\href{}{Aura Obreja$^{1,2}$}, \\
\href{}{\textbf{Nihat Oguz}$^4$},
\href{}{\textbf{William H. Oliver}$^{1,2}$},  
\href{}{\textbf{Horea-Alexandru Cărămizaru}$^5$}
\\
  $^1$Interdisciplinary Center for Scientific Computing, University of Heidelberg, D-69120 Heidelberg\\
  $^2$Zentrum für Astronomie, Institut für Theoretische Astrophysik, D-69120 Heidelberg, Germany\\
  $^3$Oxford Robotics Institute, Department of Engineering Science, University of Oxford, UK\\
  $^4$Institute for Theoretical Physics I, University of Stuttgart, D-70569 Stuttgart, Germany\\
  $^5$School of Engineering, The University of Edinburgh, Edinburgh, UK
}

\begin{document}

\maketitle

\begin{abstract}
Although integral-field spectroscopy enables spatially resolved spectral studies of galaxies, bridging particle-based simulations to observations remains slow and non-differentiable. We present RUBIX, a JAX-based pipeline that models mock integral-field unit (IFU) cubes for galaxies end-to-end and calculates gradients with respect to particle inputs. Our implementation is purely functional, sharded, and differentiable throughout. We validate the gradients against central finite differences and demonstrate gradient-based parameter estimation on controlled setups. While current experiments are limited to basic test cases, they demonstrate the feasibility of differentiable forward modelling of IFU data. This paves the way for future work scaling up to realistic galaxy cubes and enabling machine learning workflows for IFU-based inference. The source code for the RUBIX software is publicly available under \url{https://github.com/AstroAI-Lab/rubix}.
\end{abstract}

\section{Motivation}
An IFU cube is a three-dimensional dataset obtained using an integral field unit spectrograph. Two dimensions record the spatial distribution of light in the sky, while the third dimension captures the spectrum at each position. Modern IFUs such as MUSE \cite{Bacon2010} and MaNGA \cite{Sanchez2022} provide spatially resolved spectra across thousands of galaxies, offering unparalleled insight into stellar populations and dynamics. To compare these data with theoretical models, researchers typically forward model cosmological simulations into synthetic IFU cubes. Although existing software (e.g., SimSpin \cite{Harborne2020,Harborne2023}, MaNGA data utilities \cite{Ibarra-Medel2019,Sarmiento2023} or GalCraft \cite{Wang2024}) can perform this mapping, it is often CPU-bound and does not expose derivatives. This restricts their application in contemporary machine learning pipelines, where gradients are vital for optimisation, gradient-assisted simulation-based inference \cite{Holzschuh2024,Zenghal2022} and uncertainty quantification.

Differentiable programming offers a paradigm to close this gap: models are represented as computational graphs and paired with automatic differentiation to expose gradients end-to-end \cite{Blondel2024}. Frameworks like JAX \cite{jax2018github} enable scientists to build such pipelines efficiently. In astrophysics, however, a practical, end-to-end differentiable simulator for IFU cubes has so far been missing.

\texttt{RUBIX} \cite{Cakir2024} addresses this by constructing IFU cubes of simulated galaxies within a single differentiable JAX pipeline. Every stage is functional, vectorized, and designed to scale across devices. In this work we focus on two showcase examples to establish a foundation for future developments towards gradient-assisted galaxy modelling. Crucially, it demonstrates that end-to-end gradients can be obtained in IFU forward modelling, creating a starting point for subsequent integration with machine learning inference frameworks.

\section{The \texttt{RUBIX} software}
\begin{wrapfigure}[28]{l}{0.3\textwidth} 
\centering
\begin{tikzpicture}[node distance=4mm]
  \node[block] (cfg)  {config + input data};
  \node[block, below=of cfg]   (rot)  {galaxy rotation};
  \node[block, below=of rot]   (spx)  {spaxel assignment};
  \node[block, below=of spx]   (look) {spectra lookup};
  \node[block, below=of look]  (mass) {mass scaling};
  \node[block, below=of mass]  (dop)  {doppler shift};
  \node[block, below=of dop]   (wave) {wavelength resampling};
  \node[block, below=of wave]  (psf)  {point spread function};
  \node[block, below=of psf]   (lsf)  {line spread function};
  \node[block, below=of lsf]   (noise){noise};
  \node[block, below=of noise] (IFU)  {IFU cube};

  \foreach \a/\b in {cfg/rot,rot/spx,spx/look,look/mass,
                     mass/dop,dop/wave,wave/psf,psf/lsf,lsf/noise,noise/IFU}
    \draw[arrow, shorten >=1pt, shorten <=1pt] (\a) -- (\b);
\end{tikzpicture}
\caption{Different steps that happen inside the \texttt{RUBIX} pipeline. We start with a config file and input data, get a mock IFU cube and in between there are pure JAX functions.}
\label{fig:pipeline-flowchart}
\end{wrapfigure}
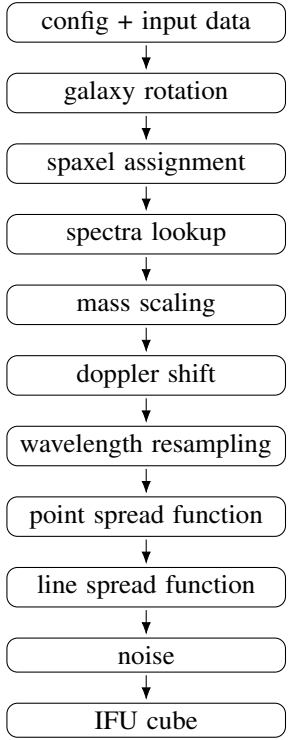

\texttt{RUBIX} is designed as a modular, purely functional pipeline that calculates IFU cubes from particle-level inputs. Its GPU accelerated capabilities were introduced in \cite{Cakir2024}. The overall flow is illustrated in Figure \ref{fig:pipeline-flowchart} as a clear sequence of stages and shows the individual physics processes to produce a realistic modelled IFU cube observation.

All components are implemented in JAX, partially vectorized via \texttt{vmap}, and parallelized across devices using \texttt{shard\_map}. This ensures end-to-end differentiability and GPU acceleration as shown in \cite{Cakir2024}, while the modular structure allows individual blocks to be modified and to be selectively chosen during forward modelling via the yaml input file.

Although tailored for astrophysical applications, \texttt{RUBIX}' modular structure with end-to-end gradient support enables direct integration with variational inference frameworks, the embedding of neural networks inside the solver, solver in-the-loop setups \cite{Um2020} and efficient coupling to simulation-based inference pipelines \cite{Cranmer2020}.

\section{Results}
\label{sect:results}
We present two illustrative examples that demonstrate that \texttt{RUBIX}' differentiability and its application to gradient-based parameter optimisation. Our model takes as input the positions and velocities of star particles with specified ages and chemical compositions (metallicities), and computes their stellar spectra using the spectral synthesis library FSPS \cite{Conroy2009} while consistently incorporating stellar dynamics.

\subsection{Gradient validation and the physical interpretation of the gradients}
\label{sect:gradientcalculation}

\begin{figure}
    \centering
    \includegraphics[width=0.95\linewidth]{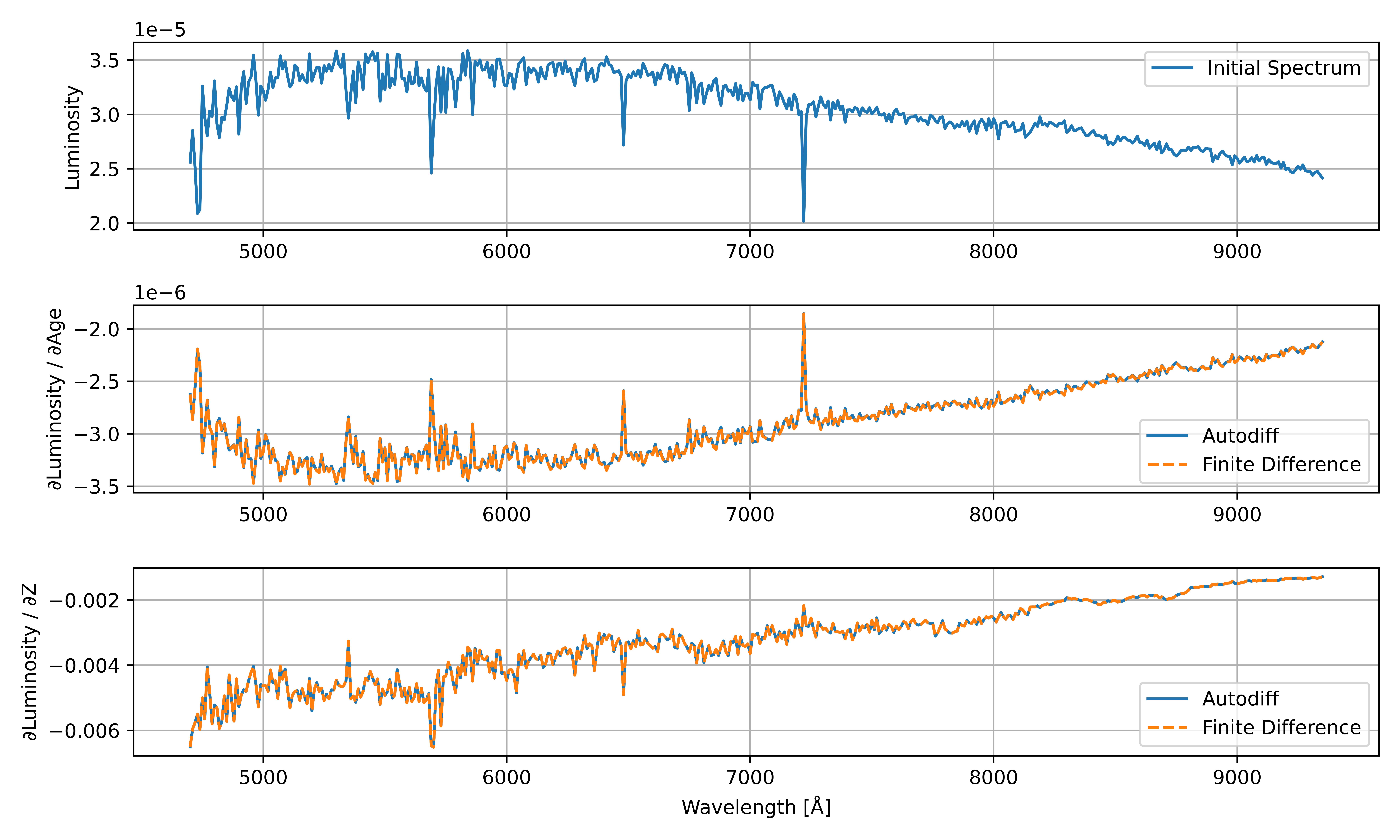}
    \vspace{-1em}
    \caption{First row shows the spectrum of a single spaxel calculated by \texttt{RUBIX}, which contains one stellar particles with age of 10 Gyrs and metallicity of ($1.4\times10^{-3}$). We calculate the gradient with respect to the stellar age in the second row and show the results for autograd (\texttt{jax.jacfwd}) and central finite difference. In the third row we show the gradient via autodiff and the finite difference with respect to the metallicity of the stellar particles.}
    \label{fig:autograd_vs_finitediff}
\end{figure}

We validate the differentiability of \texttt{RUBIX} by comparing JAX autodiff gradients with central finite differences on the spaxel spectra of a one-particle system with one spaxel\footnote{The code to reproduce these results can be found at \url{https://github.com/AstroAI-Lab/rubix/blob/eurips-anonymous/notebooks/gradient_age_metallicity_spectrum.ipynb}.}. A spaxel is a pixel with the spectral dimension.
Figure \ref{fig:autograd_vs_finitediff}
illustrates both the predicted spectrum and its parameter sensitivities. The top panel shows the luminosity generated by our model at 10 Gyrs and a metallicity of $1.4\times10^{-3}$. Prominent absorption features are clearly visible, reflecting the underlying stellar template FSPS \cite{Conroy2009}.
The middle panel compares gradients of the luminosity with respect to stellar particle age, computed via automatic differentiation (solid blue) and finite differences (dashed orange). The maximum difference in a wavelength bin between finite difference and autodiff is $4.4\times10^{-9}$ for the age and $3.9\times10^{-6}$ for the metallicity. The two approaches show excellent agreement, validating the correctness of our autodiff implementation. Negative gradients for the age dominate across most wavelengths, indicating that increasing age systematically reduces flux, particularly in the blue region where younger stellar populations contribute most strongly.
The bottom panel shows the corresponding gradients with respect to metallicity. Again, the autodiff and finite-difference methods are consistent and we find negative values across much of the optical wavelength range in agreement with physical expectations where increasing metallicity reduces flux in the blue part of the spectrum.

\subsection{Parameter estimation via gradient-based optimization}
\label{sect:adamoptimizer}
\begin{figure}
    \centering
    \includegraphics[width=0.9\linewidth]{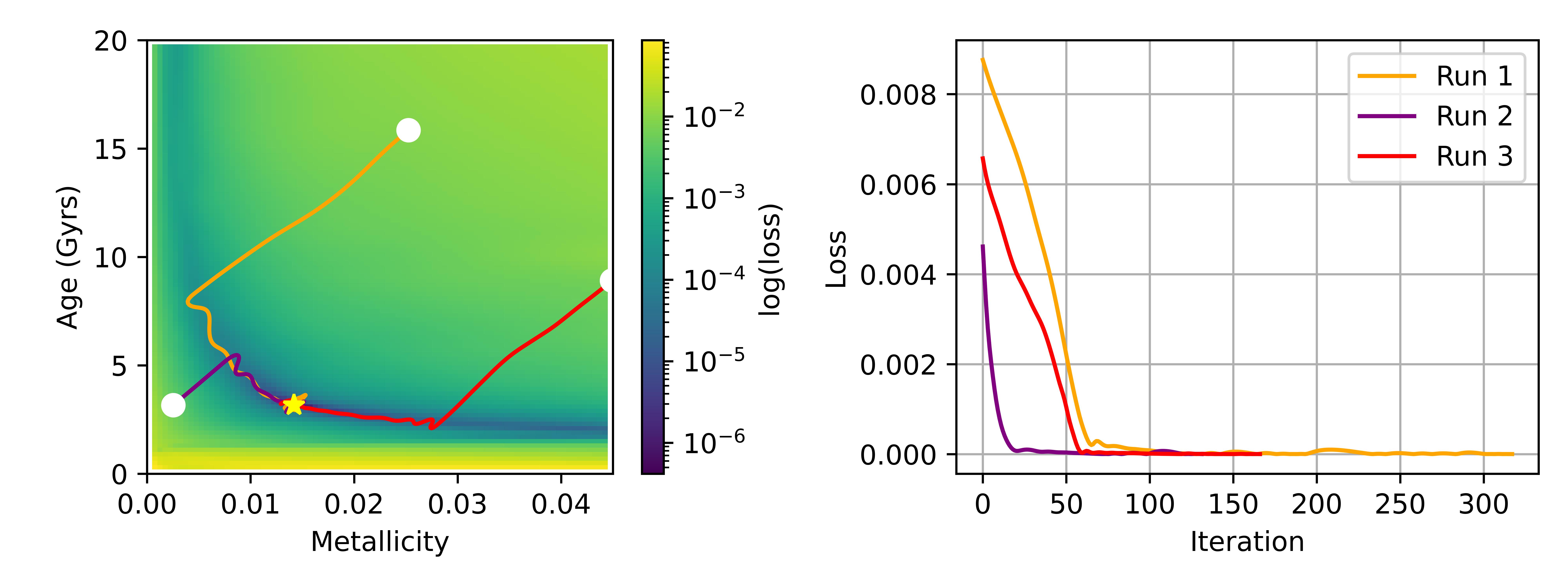}
    \vspace{-1em}
    \caption{Left: loss landscape and optimization trajectories. The true parameters are marked by the yellow star; initializations are shown as white circles. Right: loss history across iterations for three runs, showing rapid convergence.}
    \label{fig:loss}
\end{figure}

To demonstrate the utility of \texttt{RUBIX} for inference\footnote{The code to reproduce these results can be found at \url{https://github.com/AstroAI-Lab/rubix/blob/main/notebooks/gradient_age_metallicity_adamoptimizer_multi.ipynb}.}, we constructed with \texttt{RUBIX} a synthetic cube consisting of a single spaxel generated from one stellar particle with fixed stellar age ($a = 3.16$ Gyr) and metallicity ($m = 1.4\times10^{-2}$). This target cube (yellow star in Figure \ref{fig:loss}) is compared against cubes forward-modelled with \texttt{RUBIX} from arbitrary initial age and metallicity values (white circles in Figure \ref{fig:loss}). The discrepancy between target and model cube was quantified with a scalar loss $\ell$ per spaxel,
$    \ell(a, m) = \mathrm{cosine\_distance}\!\Big(\mathrm{IFU}_\mathrm{target}, \,\mathrm{IFU}_\mathrm{model}(a,m)\Big),$
using the \texttt{optax.cosine\_distance} implementation \cite{deepmind2020jax}. 
We adopted cosine distance because it provided faster convergence and less oscillation than alternative metrics tested.

Since \texttt{RUBIX} is implemented as a pure JAX pipeline, the gradient of the loss with respect to age $a$ and metallicity $m$ can be computed directly. These gradients are then used to update the input parameters with the Adam optimizer \cite{adam2017} 
(learning rate $5\times10^{-3}$, tolerance $10^{-10}$). All experiments were run on an Apple M3 CPU. The optimization converged within $\sim50$ iterations (right panel of Figure \ref{fig:loss}), yielding spectra that agree closely with the target, with residuals two orders of magnitude smaller than the flux values.

Figure \ref{fig:loss} summarizes the process. The left panel shows the log loss landscape in age–metallicity space. Optimization trajectories from different initializations rapidly descend into the L-shaped minimum region, sometimes overshooting metallicity before settling near the true parameters. The right panel shows loss histories for three runs, all decreasing steeply before flattening at low values. Oscillations around the optimum can be reduced by lowering the learning rate, at the cost of more iterations. 

\section{Discussion and limitations}
\label{sect:discussion}
The experiments presented here highlight both the opportunities and current boundaries of \texttt{RUBIX}. Our validation focused on controlled single-spaxel, one-particle setups, where both target and model come from the same forward model. These controlled setups are deliberately minimal: they provide clean benchmarks for testing differentiability and optimization, but they do not yet capture the complexity of realistic IFU cubes containing hundreds of spaxels. Extending \texttt{RUBIX} to full-galaxy simulations is technically feasible but introduces significant challenges related to memory consumption, non smooth loss landscapes and stable gradient propagation across devices.

\paragraph{Ill-posed inverse problems:} The loss landscape often contains extended valleys or L-shaped minima, which explains overshooting and oscillations in the recovered parameters. Although \texttt{RUBIX} provides accurate gradients for the synthetic example, these alone are insufficient to resolve intrinsic degeneracies. Addressing them will require more advanced optimization strategies, additional physical constraints (e.g. deep inverse priors \cite{Yang2025,Ulyanov2017}), or regularization across many spaxels.

\paragraph{Future applications:} Looking ahead, the differentiable design of \texttt{RUBIX} suggests multiple directions for expansion. Because gradients can be propagated through the full pipeline, one can in principle begin from observational data and optimize an underlying particle or distribution-function model that reproduces the observed cube. This goes beyond classical approaches based solely on forward simulations, since gradients naturally capture nonlinear relationships between model parameters and observables. Extending the current examples to realistic galaxies and testing them against survey data such as GECKOS \cite{Fraser-McKelvie2024} is a natural next step. Importantly, \texttt{RUBIX} is not limited to particle-level inference. It can be coupled with differentiable distribution function models \cite{Binney2008,Binney2012}, where optimization targets a smaller set of global parameters describing stellar distributions, kinematics, or structural properties. This reduces the dimensionality of the inference problem and provides a more interpretable connection between model parameters and astrophysical quantities of interest.

\paragraph{Uncertainty quantification via variational inference:} Beyond point estimation, the differentiable structure of \texttt{RUBIX} naturally enables uncertainty quantification. Integration with modern machine learning inference frameworks, such as simulation-based inference or variational inference, would allow posterior distributions to be approximated rather than single best-fit values. 

\section{Conclusion}
\texttt{RUBIX} demonstrates that differentiable programming can be successfully applied to challenging astrophysical forward modelling. By expressing the entire IFU modelling pipeline in JAX, we obtain end-to-end gradients and GPU-accelerated execution. Our results show that gradients are numerically consistent with finite differences and can be exploited for gradient-based parameter estimation.
The present experiments establish a very promising proof of concept for differentiable IFU cube simulation. The framework is designed with scalability in mind, and future work will address realistic galaxies and multi-spaxel cubes. 

Overall, \texttt{RUBIX} provides a foundation for combining astrophysical modelling with modern machine learning methodology. Its differentiable structure opens the path toward variational inference, solver in-the-loop setups and simulation-based inference, ultimately enabling new ML workflows for IFU data analysis.

\begin{ack}
The authors thank Kate Harborne for the support during the development of \texttt{RUBIX}. The authors thank the Scientific Software Center at Heidelberg University for the support. This work is funded by the Carl-Zeiss-Stiftung through the NEXUS program. This work was supported by the Deutsche Forschungsgemeinschaft (DFG, German Research Foundation) under Germany’s Excellence Strategy EXC 2181/1 - 390900948 (the Heidelberg STRUCTURES Excellence Cluster). We acknowledge the usage of the AI-clusters Tom and Jerry funded by the Field of Focus 2 of Heidelberg University.
\end{ack}

\bibliographystyle{plain}
\bibliography{references.bib}





\end{document}